\documentclass{ws-rv9x6}
\usepackage{subfigure}     % required only when side-by-side / subfigures are used
\usepackage{ws-rv-van}     % numbered citation/references (default)

\usepackage{graphics}
\usepackage{graphicx}

\def\m@thcombine#1#2{%
  \setbox0=\hbox{$#1$}
  \setbox1=\hbox{$#2$}
  \ifdim\wd0>\wd1
    \setbox0=\hbox to\wd1{\hss\box0\hss}
  \else
    \setbox1=\hbox to\wd0{\hss\box1\hss}
  \fi
  \mathop{\vcenter{
    \offinterlineskip\box0\box1}}}
\def\lesim{\m@thcombine<\sim}
\def\gesim{\m@thcombine>\sim}

\newcommand{\ket}[1]{\left| #1 \right\rangle}

\newcommand{\vecr}{\mbox{\boldmath $r$}}
\newcommand{\veck}{\mbox{\boldmath $k$}}
\newcommand{\vecCR}{\mbox{\boldmath $R$}}

\makeindex
\begin{document}

\chapter[ 
Spatial structure of Cooper pairs in nuclei]
{
Spatial structure of Cooper pairs in nuclei}
%Using World Scientific's Review Volume Document Style]{Using World Scientific's Review Volume\\ Document Style}\label{ra_ch1}

\author[M. Matsuo]{Masayuki Matsuo}
%\index[aindx]{Author, F.} % or \aindx{Author, F.}
%\index[aindx]{Author, S.} % or \aindx{Author, S.}

\address{Department of Physics, Niigata University,\\ 
Ikarashi Ninocho 8050, Nishi-ku, Niigata 950-2181, Japan\\
matsuo@nt.sc.niigata-u.ac.jp}

\begin{abstract}
We discuss the spatial structure of the Cooper pair in dilute neutron matter and neutron-rich nuclei 
by means of the BCS theory and the Skyrme-Hartree-Fock-Bogioliubov model, respectively.
The neutron pairing in dilute neutron matter is close to the region of the 
BCS-BEC crossover in a wide density range, giving rise to spatially compact Cooper pair
whose size is smaller than the average interaparticle distance. This behavior extends
to moderate low density ($\sim 10^{-1}$ of the saturation density) where the 
Cooper pair size becomes smallerst ($\sim 5$ fm).
The Cooper pair in finite nuclei also exhibits the spatial correlation
favoring the coupling of neutrons at small relative distances $r \lesim 3$ fm
with large probability. 
Neutron-rich nuclei having small neutron separation
energy may provide us opportunity to probe the spatial correlation since
the neutron pairing and the spatial correlation persists also in an area of 
low-density neutron distribution extending from the surface to far outside the
nucleus.  
\end{abstract}

%\markright{Customized Running Head for Odd Page} % default is chapter title.
\body

\section{Introduction}\label{Intro}

The formation and the condensation of the Cooper pairs are the essence of superconductivity 
and superfluidity in many-Fermion systems\cite{BCS}. The binding energy of the Cooper pair is closely 
related to 
the pairing gap  $\Delta$. The spatial size of the Cooper pair is 
identified to the coherence length $\xi$ of the superconductors, which
 plays important roles in
many aspects, for instance, in distinguishing
 the type I and type II
superconductors. What is the size of
the Cooper pair in the superfluidity of nuclear systems?  A simple
estimate of the coherence length $\xi$, based on the uncertainty principle
in uniform matter, leads to
$
\xi \sim \frac{\hbar v_F}{\pi \Delta}
$
with $v_F$ being the Fermi velocity\cite{BCS}.  If one considers 
saturated nuclear matter as a simplification of finite nuclei,
and adopts the typical value of the
pairing gap $\Delta \approx 12/\sqrt{A} \sim 1$ MeV appropriate
for heavy nuclei, the estimate gives $\xi \sim  20$ fm
which is much larger\cite{BM2,Brink-Broglia} than the radius of nuclei $R\approx 1.2 A^{1/3}\sim 3-7$ fm
or interparticle distance $\sim 2.5$ fm in saturated matter. However, if one considers 
extreme situations, such as dilute neutron matter and exotic nuclei with large neutron
excess, there appear new features of the 
nuclear pairing that can be related to the spatial structure of the
Cooper pair. 
It is the aim of this article to illustrate it
using a few examples.

\section{Dilute neutron matter}

The  superfluidity in neutron matter is density 
dependent\cite{TT93,Lombardo-Schulze,Dean03}.
The pairing gap can be obtained by solving the BCS equations for the bare nuclear force in
the $^{1}S$ channel at each neutron density  $\rho=k_F^3/3\pi^2$ or the Fermi
momentum $k_F$. The gap is small $\Delta \ll 1$ MeV at $k_F = 1.36$ fm$^{-1}$ 
($\rho/\rho_0=1$,  the neutron density at saturation $\rho_0=0.08$ fm$^{-3}$).
With decreasing the density  it first increases, reaching 
the maximum $\Delta \approx 3$ MeV around $k_F\approx 0.8$ fm$^{-1}$ 
($\rho/\rho_0\approx 0.2$), 
then decreases and approaches to zero
at the low-density limit. Other many-body medium effects 
 which are beyond the BCS approximation reduce
the gap, but the predictions vary depending on the
 theoretical methods\cite{Lombardo-Schulze,Cao-Lombardo,Schwenk,Fabrocini05}. Recent ab initio Monte Carlo calculations\cite{GezerlisCarlson,AbeSeki09,Gandolfi}, on the
 other hand, predict rather modest reduction by less than 50\%, and
 the qualitative features of the density dependence is kept. Having these
 reservations in mind, let us consider the structure of the 
neutron Cooper pair in the BCS approximation.\cite{Matsuo06}

The Cooper pair
wave function  can be defined, apart from the normalization, as
an expectation value of the pair operator with respect to the BCS state:
\begin{equation}
\Psi_{pair}(\vecr_1,\vecr_2) =
\left<\psi(\vecr_1\uparrow)\psi(\vecr_2\downarrow)\right>
=\sum_{\veck} u_k v_k e^{i\veck\cdot\vecr}.
\label{eq.matter}
\end{equation}
It is a function of the relative coordinate $\vecr=\vecr_2-\vecr_1$
of the two neutrons,  and in the momentum space it 
is a product of the $u$ and $v$ factors. 
Examples of the Cooper pair wave functions are shown in Fig. 1
 for two different densities\cite{Matsuo06}. The wave function exhibits
an oscillatory behavior characterized by the Fermi wave length $2\pi/k_F$ and
an overall decay profile whose asymptotic form is
exponential $\sim \exp(-r\Delta/\hbar v_F)$  (
for large relative distance $r=|\vecr_2- \vecr_1|$) whose length scale
is nothing but the coherence length, or the size of the Cooper paper\cite{BCS}. 
More precisely, the coherence length can be calculated as the rms radius of the
Cooper pair
$\xi =\sqrt{\left<r^2 \right>}$ with 
$\left<r^2 \right>=\int d\vecr r^2 |\Psi_{pair}(\vecr)|^2/\int d\vecr |\Psi_{pair}(\vecr)|^2$.

An interesting feature of the neutron Cooper pair in superfluid neutron matter is that its size
also varies significantly with changing the neutron density (See Fig.1(a)). From a very large value
$\xi=46$ fm at $\rho/\rho_0=1$, the coherence length $\xi$ decreases sharply with decreasing
the density. The coherence length takes the smallest values $\xi=5-8$ fm for a rather
wide range of the density $\rho/\rho_0=0.2-10^{-2}$, and it increases gradually with decreasing
the density. 

\begin{figure}
%\centerline{\psfig{file=matter_3.eps,angle=270,width=9cm}}
\centerline{\psfig{file=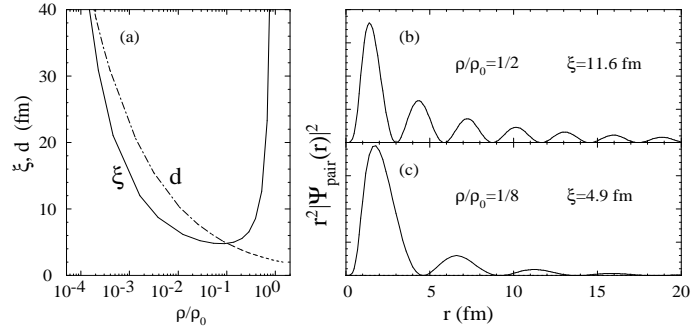,angle=270,width=9cm}}
\caption{(a) The coherence length $\xi$ and the average interparticle distance $d$ in
superfluid neutron matter, plotted as a function of the neutron density $\rho/\rho_0$
(with $\rho_0=0.08$ fm$^{-3}$). (b)(c) The Cooper pair wave function 
$r^2|\Psi_{pair}(r)|^2$ at densities $\rho/\rho_0=1$ and $1/8$.
The bare force G3RS\cite{Tamagaki} is used in the present BCS calculation.\cite{Matsuo06}
}\label{fig1}
\end{figure}

The Cooper pair wave function at densities where the coherence length is the smallest
is very different from that of of the electron Cooper pair in the traditional metal superconductors.
An example is shown in Fig.1 (c), which is for $\rho/\rho_0=1/8$ where
the coherence length $\xi=4.9$ fm is close to the minimum value.
It is seen that 
the oscillatory behavior 
is strongly suppressed. The probability distribution is
concentrated ($\sim 80\%$) at small relative distances 
within the first node $r < \pi/k_F\approx 4.5$ fm, and the probability at the second
and third bumps is very small. 
This is because 
the size of the Cooper pair ($\xi=4.9$ fm) is almost equal to the position of the first node
$\pi/k_F$ which is nothing but 
 the average interparticle distance
$d =\rho^{-1/3}\approx \pi/k_F$. The size of the Cooper pair is "small" in this sense.
This is quite contrasting to the metal superconducters where the Cooper pair size $\xi$
is thousands times larger than the average interparticle distance $d$. 
The situation of the "small" Cooper pair $\xi \lesim d$ is seen in a wide interval of
densities $\rho/\rho_0 =10^{-4} - 10^{-1}$ (Fig.1(a)). The shape of the Cooper
pair wave function at these densities is similar to that of Fig.1(c), and the probability is
 even more concentrated in the first bump although the absolute size is larger
at very low densities $\rho/\rho_0 \approx 10^{-4} - 10^{-2}$. 
It is noted that the Cooper pair at moderate low densities $\rho/\rho_0 \approx 10^{-1} - 0.5$
exhibits also the strong spatial correlation at small relative distances.
The wave function at $\rho/\rho_0=0.5$ is shown in Fig.1(b). In this case
 the calculated coherence length $\xi=11$ fm is
a few times larger than the average interparticle distance $d=2.8$ fm.
 Nevertheless the concentration of the 
probability within the relative distance $r \lesim 3$ fm (in the first bump) is
signfinicant, and the probability in $r < 3$ fm 
reaches as large as $\sim 50\%$.  

The situation of the 
 small Cooper pair $\xi/d \lesim 1$ is related to the so-called BCS-BEC
crossover phenomenon\cite{Leggett,Nozieres,Melo,Engelbrecht,Randeria}, 
which has been 
discussed intensively in ultra-cold Fermi atom gas
in a trap\cite{Regal,review-cold-gas}. It is a phenomenon which can occur generally in any kind of many-Fermion
superfuluid systems by changing the strength of the interparticle attractive force or the density.
In a situation of the weak interaction,  which the original BCS theory has dealt with,
the bound pair (the Cooper pair) can be formed only in the medium. 
However, if the interaction is as strong as to form a bound
pair (a composite boson) even in the free space, the condensed phase is more 
close to a condensate of the composite bosons, i.e. the Bose-Einstein condensate (BEC). 
 The BCS-BEC crossover is characterized by
the ratio $\xi/d$ of the coherence length and the average interparticle distance
and the ratio $\Delta/e_F$ of the pairing gap and the Fermi energy.
The weak-coupling BCS and the BEC limits correspond to $\xi/d \gg 1, \ \Delta/e_F \ll 1$
and $\xi/d \ll 1,\ \Delta/e_F \gg 1$, respectively while the region of
the crossover may be related to $0.2 \lesim \xi/d \lesim 1.2$ and $0.2 \lesim \Delta/e_F \lesim 1.3$.
\cite{Melo,Engelbrecht,Randeria}
At the midway of the crossover, called the
unitarity limit, the interaction strength is on the threshold to
form the isolated two-particle bound state, and the values are $\xi/d = 0.36, \Delta/e_F = 0.69$. 
In the BCS calculation discussed above\cite{Matsuo06}, small $\xi/d$ ratio $0.7 - 1.2$ and large
$\Delta/e_F$ ratio $0.2-0.4$ is realized at $\rho/\rho_0 \sim 10^{-4}-10^{-1}$.
(Note that also in an ab initio calculation\cite{GezerlisCarlson}, the large gap ratio
$\Delta/e_F\sim 0.2-0.3$ is obtained in approximately the same
but slightly small density region.) We can regard dilute neutron matter in
the wide low-density interval  
$\rho/\rho_0=10^{-4}- 10^{-1}$ (or in slightly narrower interval)
as being in the crossover region. We note here that the nuclear force in the $^{1}S$ channel
has a large scattering length $a=-18$ fm,
indicating that the interaction strength is very close to the threshold to
form a two-neutron bound state. The small Cooper pair
$\xi/d \lesim 1$ at low densities originates from the nature of the nuclear force.

\section{Cooper pair in neutron-rich nuclei}

Let us consider the spatial structure of the Cooper pair  in
finite nuclei.

The spatial structure of the correlated two neutrons has been discussed
intensively
for two neutrons in the light two-neutron halo nuclei $^{11}$Li and $^{6}$He
 in  (inert or active) core plus two neutron models
\cite{Esbensen,Hansen,Ikeda,Zhukov,Barranco01,Hagino05,Hagino07,HaginoIOP,Myo08}. 
A common prediction is 
that the valence halo neutrons exhibit a spatial
correlation favoring the 'di-neutron' configuration with two neutrons coupled at small relative 
distances. The spatial correlation is also discussed in stable heavy nuclei with
closed-shell core plus two neutrons, e.g.
$^{206,210}$Pb, by means of shell model approaches.\cite{Bertsch,Ibarra,Janouch,Catara84,Ferreira}
One can generalize these findings by using the Hartree-Fock-Bogoliubov (HFB) method,
which can be applied to a wide class of open shell nuclei including isotopes
very close to the drip-line and also to non-uniform matter.

Let us start defining the wave function of the Cooper pair in finite nuclei. It may be given by  
\begin{equation}
\Psi_{pair}(\vecr_1,\vecr_2) =
\left<\Phi_{A-2}|\psi(\vecr_1\uparrow)\psi(\vecr_2\downarrow)|\Phi_{A}\right>
\label{eq.finite}
\end{equation}
using the pair correlated ground states $\Phi_{A}$ and $\Phi_{A-2}$. This 
represents the probability amplitude of removing two neutrons (positioned at
 $\vecr_1$ and $\vecr_2$) from the ground state $\Phi_{A}$, and
leaving the remaining system in the ground state  $\Phi_{A-2}$. 
Provided that the  ground state is described
within the HFB framework, where the ground states with different
nucleon numbers are represented by a single HFB state $\Phi_{{\rm HFB}}$,
the definition  Eq.(\ref{eq.finite}) can be replaced with the expectation value
as in Eq.(\ref{eq.matter}). Then, since the HFB state
is a generalized Slater determinant consisting of the Bogolviubov 
quasiparticle states, this quantity is evaluated\cite{MMS05}
as a sum over all quasiparticle states $i$
\begin{equation}
\Psi_{pair}(\vecr_1,\vecr_2)=
\left<\Phi_{{\rm HFB}}|\psi(\vecr_1\uparrow)\psi(\vecr_2\downarrow)|\Phi_{{\rm HFB}}\right>
=\sum_i \varphi_i^{(1)}(\vecr_1\uparrow)\varphi_i^{(2)*}(\vecr_2\downarrow)
%\approx \sum_i u_i v_i \phi_i((\vecr_1\uparrow)\phi_i(\vecr_2\downarrow)
\label{eq.finite2}
\end{equation} 
using the first and the second components of
the quasiparticle wave function
$\phi_i(\vecr\sigma)=(\varphi_i^{(1)}(\vecr\sigma),\varphi_i^{(2)}(\vecr\sigma))$. 
In the following we show the results of our HFB calculation, which adopts
the Skryme functional and the density-dependent contact 
interaction as a phenomenological pairing force\cite{Matsuo07,Matsuo10}.  The parameter set of the
pairing interaction is such that it reproduces the scattering length $a=-18$ fm
in the low-density limit, and reproduces the average pairing gap in known nuclei\cite{Matsuo07,Matsuo10}. 

\begin{figure}
\centerline{
%\psfig{file=Sn142_450_ifkei1_surface_mod.eps,angle=270,width=5.5cm}
%\hspace{2mm}
%\psfig{file=jlcut.rpao.142Sn.7fm.3.eps,angle=270,width=4.5cm}
%\psfig{file=fig2L.eps,angle=270,width=5.8cm}
\psfig{file=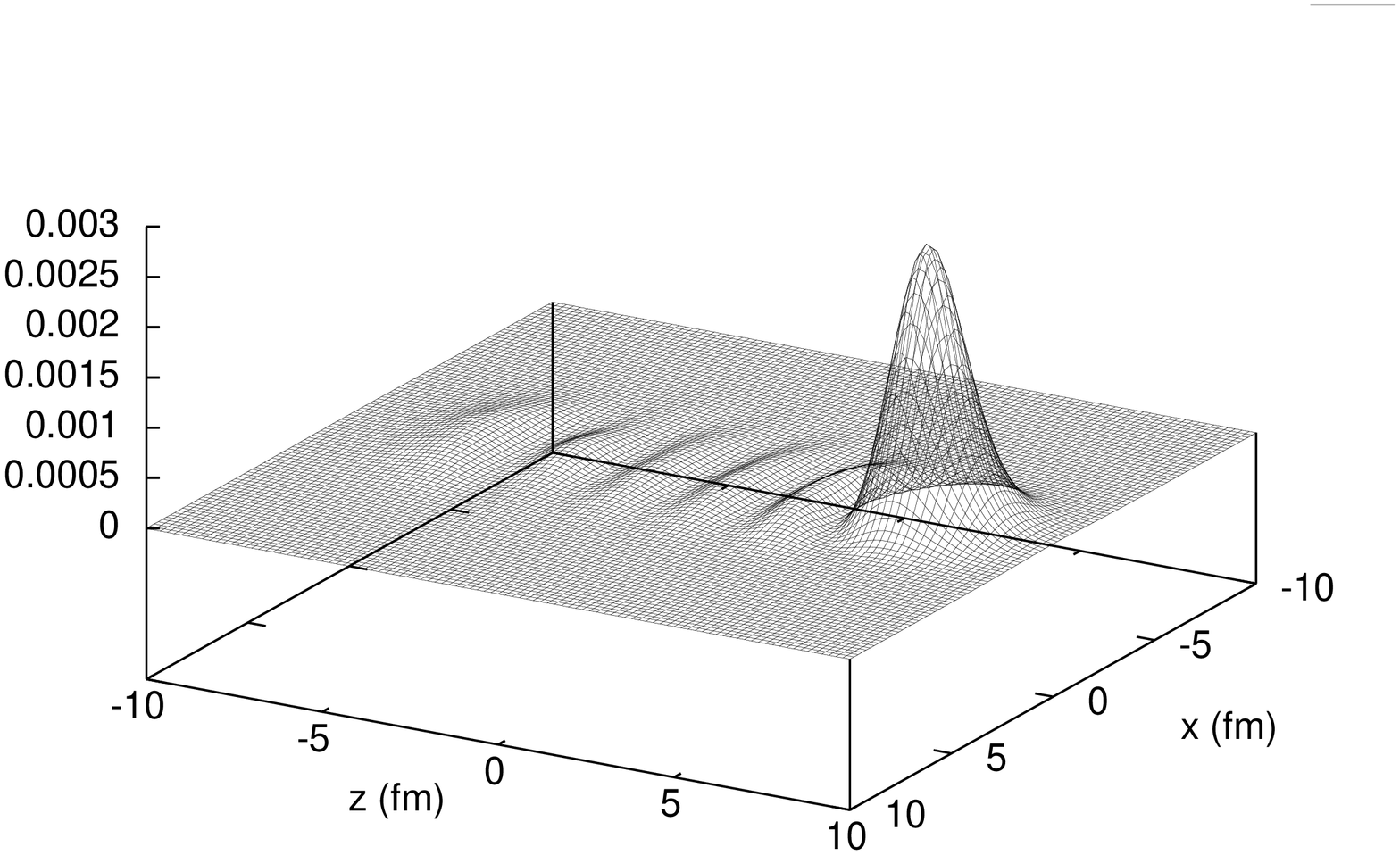,angle=270,width=5.8cm}
\hspace{2mm}
\psfig{file=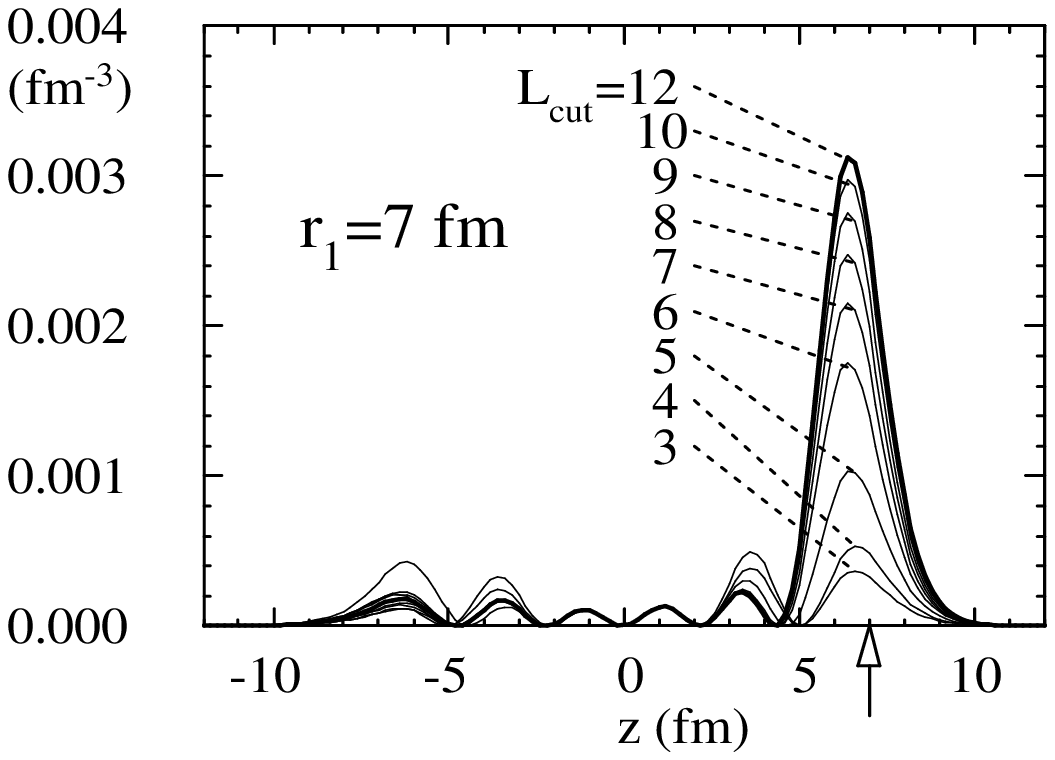,angle=270,width=4.5cm}
}
\caption{(Left) Cooper pair wave function $|\Psi_{pair}(\vecr_1,\vecr_2)|^2/\rho_n(\vecr_1)$ 
in neutron-rich nucleus
$^{142}$Sn, plotted as a function of $\vecr_2$ on the $xz$ plane while the coordinate
$\vecr_1$ is fixed to $(0,0,7)$ fm located slightly outside the surface.
(Right) The same but plotted along the $z$-axis. Different curves are results
obtained by putting cut-off's with respect to the orbital angular momentum $l$
of the single-particle orbits. 
}
\label{fig2}
\end{figure}

An example calculated for $^{142}$Sn is shown in Fig.2. 
Here one neutron is
fixed at the position slightly outside the nucleus $r_1=7$ fm
and the probability distribution $|\Psi_{pair}(\vecr_1,\vecr_2)|^2$ is 
plotted as a function of $\vecr_2$. It shows that 
the second neutron has a large probability ($\sim 50\%$) to be correlated at small 
relative distances 
$|\vecr_1-\vecr_2|\lesim 3$ fm to the partner neutron.
The spatial correlation seen here is generic in a sense that it is seen 
systematically in Ca, Ni, and Sn isotopes including both stable and neutron-rich nuclei\cite{MMS05}.
The strong spatial correlation is also seen in other HFB calculations which adopt the
finite-range Gogny force as the effective pairing force.\cite{Pillet07,Pillet10} 

I emphasize here that a large single-particle space is necessary  
in describing the spatial correlation\cite{MMS05}. 
In order to describe the correlation with the length scale
 $D\sim 3$ fm, the single-particle basis needs to cover a momentum range up to
$p_{max} \sim h/D$, which corresponds to a maximal energy 
$e_{max} \sim p_{max}^2/2m \sim 80$ MeV, or a maximal angular momentum
$l_{max} \sim Rp_{max} \sim 10 \hbar$ (for the nuclear radius $R\sim 5$ fm). 
This is demonstrated  in Fig.2(right), where the summation over the quasiparticle
states $i$ in Eq.(\ref{eq.finite2}) is truncated by introducing a cut-off with respect to
the orbital angular momentum $l$.  Single-particle orbits with large angular
momentum up to $l_{max} \sim 10$ have sizable contributions.  Note that 
in $^{142}$Sn with  $N=92$ the Fermi energy is around the
$3p_{3/2}$ orbit, and the maximal orbital angular momentum of the orbits
occupied in the independent particle limit is $l=5$. The single-particle states 
with $l=5-10$ lie high above
the Fermi energy. 
If one uses the harmonic
oscillator basis, it should include $\sim$ 10 oscillator quanta. In fact, all the
HFB calculations\cite{MMS05,Pillet07,Pillet10} where the strong spatial correlation in the Cooper pair
wave functions is demonstrated adopt such a large single-particle space.

Equivalently, a small single-particle space is
insufficient.  If
we restrict ourselves to a single-$j$ shell $(nlj)$, i.e., 
the sum in Eq.(\ref{eq.finite2}) is restricted to the magnetic substates of the orbit $(nlj)$,
we obtain the angular correlation\cite{Mottelson,Ring-Schuck} $P_l(\theta_{12})$
for small relative angles $\theta_{12}\lesim 1/l$, but the correlation with respect to
the radial direction is not produced.  Inclusion of all the orbits
in one oscillator shell still has deficiency\cite{Pillet07,Pillet10}. The Cooper pair wave function in 
this case exhibits an artificial symmetry
$\Psi_{pair}(\vecr_1,\vecr_2)=\pm \Psi_{pair}(-\vecr_1,\vecr_2)$ because of
the common single-particle parity, and the probability appears not only 
around $\vecr_2 \sim \vecr_1$, but also around the mirror reflected
position $\vecr_2 \sim -\vecr_1$.

\begin{figure}
\centerline{
%\psfig{file=fig.sn142.eps,angle=270,width=6cm}
%\hspace{2mm}
%\psfig{file=rpa.sn142.sn120.9fm.2.eps,angle=270,width=4.9cm}
\psfig{file=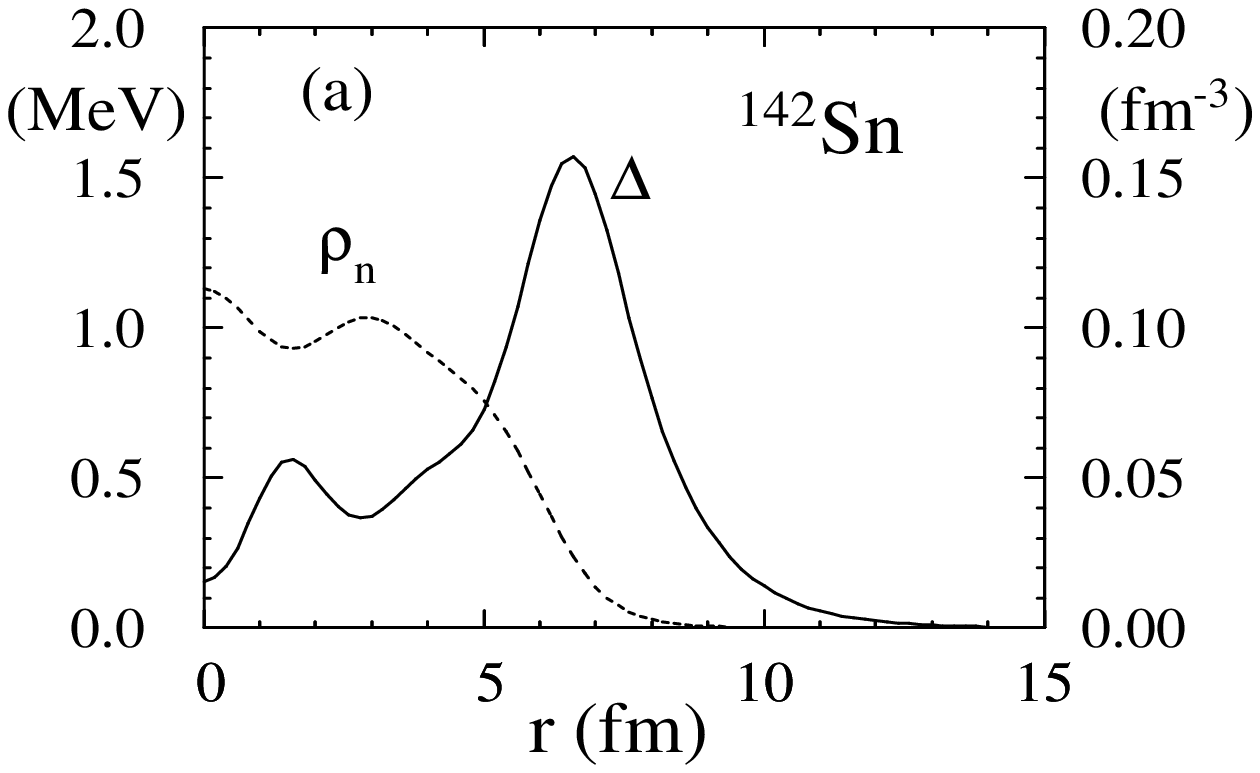,angle=270,width=6cm}
\hspace{2mm}
\psfig{file=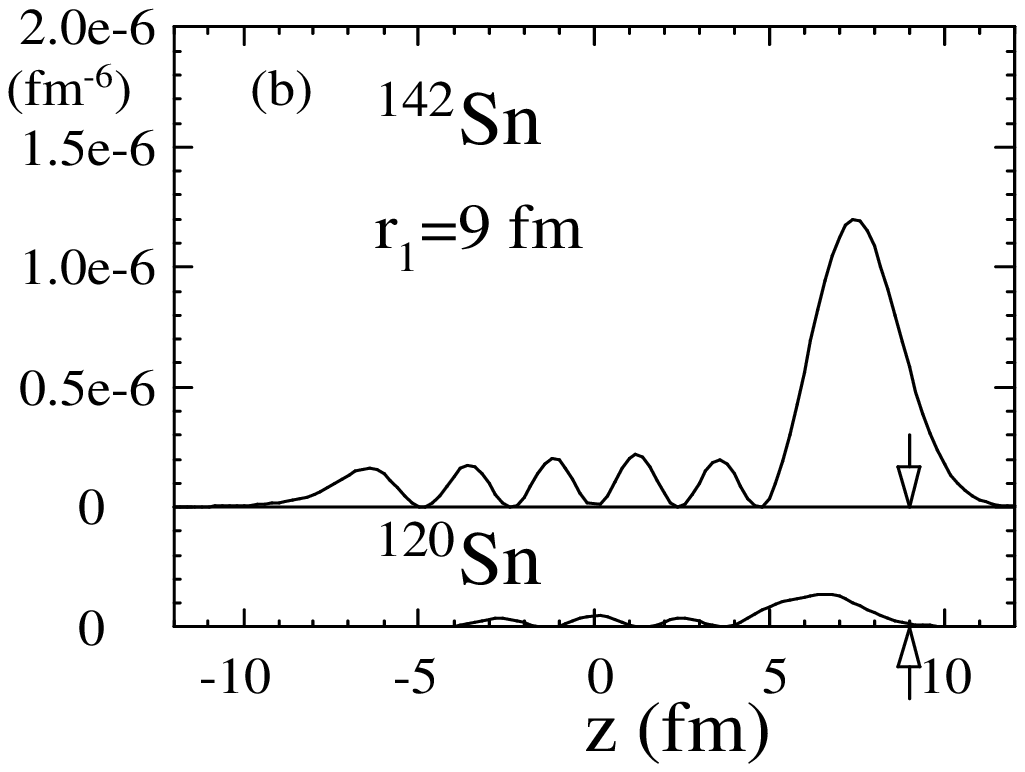,angle=270,width=4.9cm}
}
\caption{(a) Pair potential $\Delta(r)$ in
$^{142}$Sn obtained in the Skyrme-HFB model\cite{Matsuo07,Matsuo10} (solid curve),
and the neutron density $\rho_n(r)$ 
(dotted curve). (b) The Cooper pair wave function $|\Psi_{pair}(\vecr_1,\vecr_2)|^2$
in $^{142}$Sn and $^{120}$Sn, with the first coordinate fixed at $\vecr_1 = (0,0,9)$ fm and plotted
as a function of $\vecr_2$ along the z-axis.
}
\label{fig3}
\end{figure}

The spatial correlation of neutron Cooper pairs plays an important role if
we consider neutron-rich nuclei with small neutron separation energy. Nuclei of 
this kind often accompany low-density distribution of neutrons, called skin or 
halo, extending from the nuclear surface toward the outside. 
Figure 3(a) is an example of the pair potential $\Delta(r)$   for the very neutron rich
nucleus $^{142}$Sn  obtained in the same Skyrme-HFB calculation as in Fig.2.  The 
pair potential $\Delta(r)$
exhibits significant enhancement around $r\sim 5-8$ fm, which is slightly
outside the nuclear surface (the corresponding neutron density there is
about 1/2-1/10 of the central density). The pair potential decreases rather
slowly  with moving outside the surface region,
and it is about to diminish only at very large distances $r\gesim 12$ fm.
It is much more extended than the neutron density. Furthermore
the spatial correlation persists in this low density region as shown
in Fig.3(b). We note that the spatial correlation is present also in
stable open shell nuclei\cite{MMS05,Pillet07,Pillet10}, and it is enhanced around
the nuclear surface. However the nucleons (and hence the Cooper pairs) do not
penetrate far outside the surface in stable isotopes (cf bottom panel of Fig.3(b)). 
The pair correlations
in the dilute surrounding is a unique feature of weakly bound nuclei.

\section{Probing the spatially correlated Cooper pair}

\subsection{Soft modes}

If spatially correlated di-neutrons exist in nuclei, especially in the low-density skin/halo
region, there may emerge new modes of excitation
reflecting the motion of di-neutron(s). This simple idea\cite{Hansen,Ikeda} has been a focus of theoretical
and experimental studies of the soft dipole excitation in two-neutron
halo nuclei. Although the reality is not that simple, 
the  core+n+n models\cite{Hagino05,Myo08,Barranco01} of $^{11}$Li explain
the observed large E1 strength of the soft dipole excitation\cite{Nakamura}
in terms of the pairing and the spatial correlation of the valence halo neutrons.
It is interesting to explore possibility of similar excitation modes in heavier mass neutron-rich
nuclei, where more than two weakly bound neutrons contribute to the pair correlation.

A useful scheme to describe excitation modes built on the
pair correlated ground state is the quasiparticle random phase approximation (QRPA).
Let us take the formulation based on the same Skyrme-HFB model that is used for the
description of the ground state.\cite{Serizawa09,Matsuo10,Shimoyama11} Having a QRPA
excited state $\ket{n,LM}$, one can calculate the two-particle amplitude 
$\left< n, LM | \psi^\dagger(\vecr_1\uparrow)\psi^\dagger(\vecr_2\downarrow) | 0_{gs} \right>$
which tells us how two particles move in the excited state $\left| n ,LM\right>$ 
in reference to the ground state (of the $N-2$ system). For simplicity
let us look at the zero-range part at $\vecr_1=\vecr_2$ of the amplitude:
\begin{equation}
P_n^{pair}(\vecr) = \left< n, LM | \psi^\dagger(\vecr\uparrow)\psi^\dagger(\vecr\downarrow) | 0_{gs} \right>,
\end{equation} 
which is called the pair transition density.

\begin{figure}
\centerline{
\psfig{file=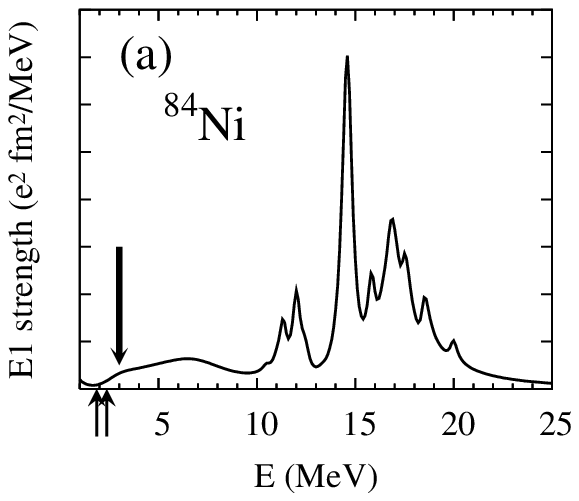,angle=0,width=5cm}
\psfig{file=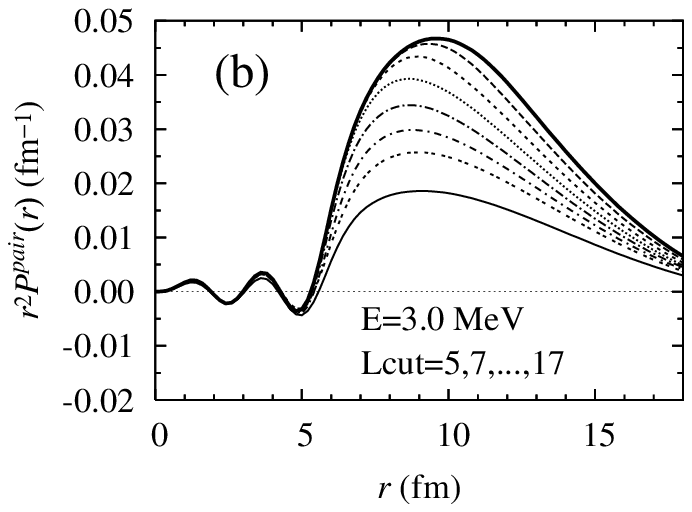,angle=0,width=6cm}
}
\caption{(a) The $B(E1)$ strength function in neutron-rich nucleus $^{84}$Ni,
obtained with the Skyrme-HFB + continuum QRPA method.\cite{Serizawa09} The large strengths around
$E=10-20$ MeV are the giant dipole resonance (GDR) while some amount of strength
is distributed just above the one- and two-neutron separation threshold energies
(the small arrows). (b) The neutron pair transition density $P^{pair}(r)$ 
of the soft dipole mode
evaluated at $E=3$ MeV (marked with the big arrow in (a)). Figures taken from Ref.\cite{Serizawa09}.
}
\label{fig4}
\end{figure}

Figure 4 is an example of soft dipole excitation 
which suggests motion of the spatially correlated di-neutrons\cite{Serizawa09}.
The soft dipole excitation is seen here as a bump of the E1 strength which
lies just above the neutron separation energies 
($S_{1n},S_{2n}=1.9,2.4$ MeV). In neutron-rich Ni isotopes beyond the $N=50$ shell
closure
both of the one- and the two-neutron separation energies are calculated to  be very low 
$S_{n},S_{2n} \approx 1-3$ MeV. In such weakly bound nuclei, the low-lying 
dipole modes appear just above the separation energy since
it is possible to excite a bound neutron to unbound orbits in the continuum,
letting the neutron escape from the nucleus. If the pair correlation is
taken into account, however, the mode is dominated by the pair motion
rather than by a simple particle-hole (or independent two-quasiparticle excitation).
Consequently the pair transition density 
$P_n^{pair}(\vecr)$ has larger amplitude, especially for $r > R_{surf}$, as
seen in Fig.4(b).
It is not explicit in this figure whether the neutron pair in the excited state is 
spatially correlated, but we can infer it from the observation that 
a large number of orbital angular momenta $l$ reaching more than 10$\hbar$ 
have significant and coherent contributions to the pair transition density. As we discussed
above (cf. Fig.2), large $l$ implies a spatial correlation at small distances 
between the two neutrons.

A similar mode of excitation having the character of di-neutron motion is 
predicted also in the octupole response in the same istopes $^{>80}$Ni beyond $N=50$.\cite{Serizawa09} 
It is a smooth distribution of neutron strength 
lying just above the threshold energy (like the soft dipole mode), and it coexists 
 from 
the octupole surface vibrational mode of the isoscalar character seen in many of
stable nuclei. 
 
In contrast to the light two-neutron halo nuclei, the presence of the spatial correlation does not
influence strongly the E1 strength of soft dipole excitation 
in heavy neutron-rich nuclei such as $^{84}$Ni. We need 
other probes which are directly connected to the pair transition density. 
Since the soft dipole excitation in $^{11}$Li and
in $^{>80}$Ni is located above the two-neutron separation energy, one can expect that
momentum distribution/correlation of
two neutrons  emitted from the
soft mode may carry information on the spatial correlation of the neutron pair.
Quantitative theoretical description of the two-neutron correlation is achieved only for the
core+n+n models\cite{HaginoNN,KikuchiNN} for $^{11}$Li and $^{6}$He, and experimental information
is very scarce so far\cite{ChulkovNN,NakamuraNN}. It is possible to describe the 
two-neutron correlation in the continuum also in the framework of the QRPA since 
the information on the directions of two neutrons are contained
in the pair transition density 
$\left< n, LM | \psi^\dagger(\vecr_1\uparrow)\psi^\dagger(\vecr_2\downarrow) | 0_{gs} \right>$
especially in its asymptotic form at $|\vecr_1|,|\vecr_2| \rightarrow \infty$.
It is an interesting future subject to study
in heavier neutron-rich nuclei such as $^{>80}$Ni using the HFB+QRPA formalism.

\subsection{Two-neutron transfer}

The two-neutron transfer reactions such as (p,t) 
and (t,p) are known as a good probe 
to the pair correlation in the ground state\cite{Yoshida62,Broglia73,BM2,Brink-Broglia}. More precisely
it can be regarded as a probe of the Cooper pair
wave function, especially its behavior at small relative distances
between the paired neutrons.  Consider the (p,t) reaction 
populating  the ground state of the neighboring
$N-2$ nucleus in the single-step DWBA and the zero-range approximation. Then 
the transition matrix elements involves the form factor\cite{Broglia73,Glendenning} 
$$
F(\vecCR)=\int d\vecr
\left<0_{gs,N-2}|\psi(\vecCR+\vecr/2\uparrow)\psi(\vecCR-\vecr/2\downarrow)|0_{gs,N}\right>\phi(\vecr)
$$
which is the convolution of  
the Cooper pair wave function $\Psi_{pair}(\vecCR+\vecr/2,\vecCR-\vecr/2)$  
with the two-particle wave function $\phi(\vecr)$ in the triton.
Noting the small radius of the triton  $\sim 2$ fm, we see immediately 
the form factor picks up the correlation
at small relative distances in the Cooper pair wave functions.
It is then not a very bad approximation to utilize
the Cooper pair wave function at zero relative distance $\vecr=0$, i.e.
$\Psi_{pair}(\vecCR,\vecCR)=\left<\psi(\vecCR\uparrow)\psi(\vecCR\downarrow)\right>\equiv P_{pair}(\vecCR)$
as a substitute of the form factor assuming $F(\vecCR)\propto P_{pair}(\vecCR)$.
$P_{pair}(\vecCR)$ is nothing but the pair density $\tilde{\rho}(\vecCR)$ implemented automatically in 
the Skyrme-HFB model using the pairing force of the contact type\cite{DobHFB2}. 

\begin{figure}
\centerline{
\psfig{file=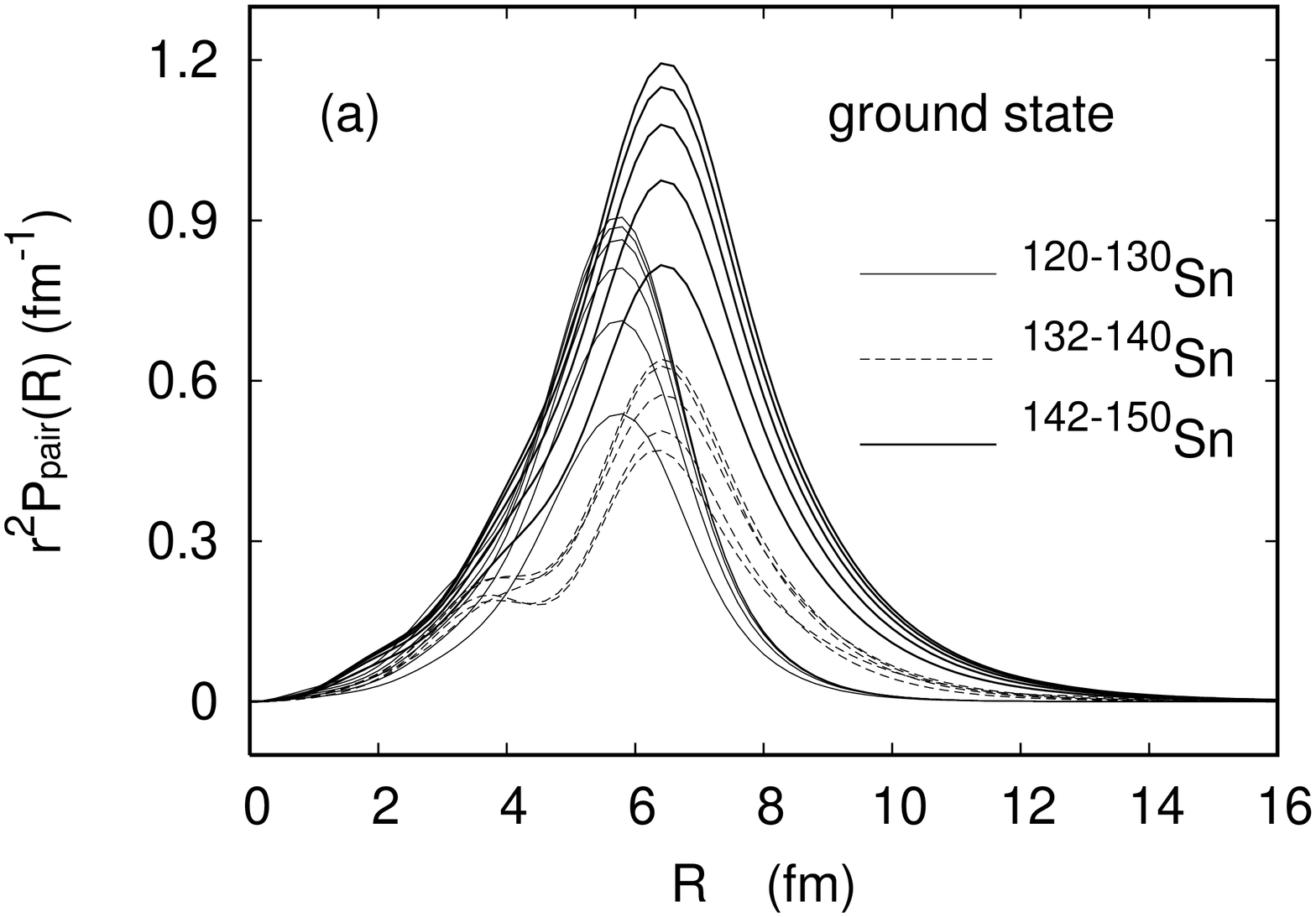,angle=270,width=5.5cm}
\psfig{file=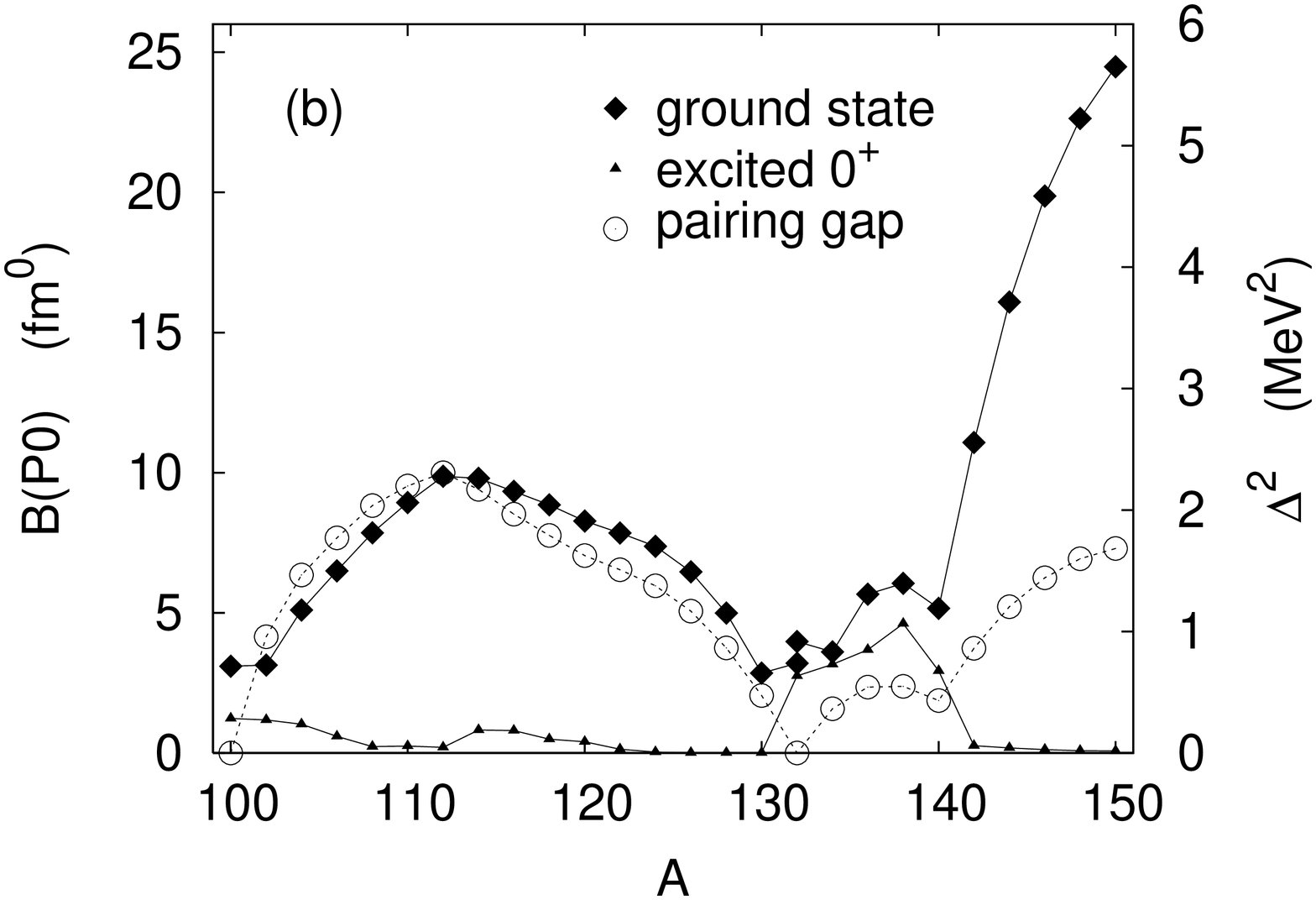,angle=270,width=5.5cm}
}
\caption{(a) Neutron pair transition density $P_{pair}(\vecCR)$ for the ground state transition,
evaluated for even-even Sn isotopes for $A=120-130$(thin solid curve),
$A=134-140$ (dotted), and $A=142-150$ (thick solid). 
(b) The two-neutron transfer strength $B(P0)$ for the ground state transition (filled diamond),
the strength for two-neutron addition transfer for the excited $0^+$ state (small triangle),
and the squared pairing gap $\Delta^2$ (open circle)
for the even-even Sn isotopes. The horizontal axis is the mass number $A$.
}
\label{fig5}
\end{figure}

An example\cite{Shimoyama11} of the calculated pair transition density $P_{pair}(\vecCR)$ is shown in Fig.5(a)
for Sn isotopes covering from stable isotopes to very neutron-rich $^{150}$Sn.
It is seen that the radial dependence of 
 $P_{pair}(\vecCR)$ suddenly changes at the $N=82$ shell closure (at $^{132}$Sn).
In neutron-rich isotopes beyond $N=82$, the amplitude extends far outside the nuclear 
surface  $r > R_{surf}+ 3$ fm ($\gesim 9$ fm). This happens because neutron single-particle
orbits above the $N=82$ shell gap are bound  only weakly, and the weakly 
bound neutrons have density distributions extended far outside the nuclear surface.
 (The one-neutron
separation energy is the order of $\sim 2-3$ MeV for $A>132$, while it is more than
8 MeV in isotopes with $A \leq 132$.)
Consequently both the pair potential $\Delta(r)$ and the Cooper pair
wave function keep non-negligible magnitude  even far outside (See also Fig.3).  
As seen in the figure the amplitude  $P_{pair}(\vecCR)$ extends up to 
$r \sim 12$ fm for the isotopes $A>140$.

The above observation leads to an expectation that the (p,t) and (t,p) cross sections may 
be enhanced considerably as the neutron separation energy becomes small\cite{DobHFB2}.
An estimate of the isotopic trend, much simpler than the DWBA calculation,
is shown in Fig.5(b). Here is plotted  the  
'strength' which is defined by
$
B(P0)
=\left|
\int d\vecCR P_{pair}(\vecCR)\right|^2.
$
It is illuminating to compare it with the isotopic trends of
 the pairing gap $\Delta$ squared ($\Delta$ being an average value of
the pair potential $\Delta(r)$). If the pair potential and the pair
transition density are confined in the nuclear volume, a proportionality relation
$B(P0) \propto \Delta^2$ is expected\cite{Yoshida62} in analogy with the 
$B(E2)$ of the deformed rotor since the pair gap is a deformation parameter\cite{Broglia73,BM2}.
We see in Fig.5(b) that the proportionality $B(P0) \propto \Delta^2$ valid
for $100<A<132$ is violated
for $A>132$ and especially $A>140$, where the strength $B(P0)$ significantly
increases. 
The two-neutron transfer reaction on the neutron-rich isotopes (e.g.
the Sn isotopes with $A>132$) thus provides us a tool to probe the Cooper pair
wave function in the low-density region far outside the nuclear surface.
It is predicted\cite{Shimoyama11} also that the isotopes $^{134-140}$Sn with $A=134-140$ exhibit
a precursor phenomenon, i.e., an anomalously large two-neutron transfer strength 
of (t,p) type for the transitions to the excited $0^+$ states (Fig.5(b)).
Recently two-neutron transfer experiment on the halo nucleus $^{11}$Li has become available, 
and the crucial role of the pair correlation is demonstrated.\cite{Tanihata08,Potel} 
We wait for a future experiment using the radioactive beams of neutron-rich
Sn isotopes with $A>132$.

This work was supported by
the Grant-in-Aid for Scientific Research 
(Nos. 21105507, 21340073 and 23540294) from the Japan
 Society for the Promotion of Science.


\begin{thebibliography}{9}

%\input{mybibws.tex}

\bibitem{BCS} 
J.~Bardeen, L.~N.~Cooper, and 
J.~R.~Schrieffer, \emph{Phys. Rev.} {\bf 108}, 1175 (1957).

P.~G.~de~Gennes, \emph{ Superconductivity of Metals and Alloys},
Benjamin (1966).

M.~Tinkham, \emph{ Introduction to Superconductivity},
McGraw-Hill (1975).

\bibitem{BM2}
A.~Bohr and B.~R.~Mottelson, \emph{ Nuclear
Structure}, vol. II,  Benjamin (1975).

\bibitem{Brink-Broglia}
D.~M.~Brink and R.~A.~Broglia, \emph
{ Nuclear Superfluidity: Pairing in Finite Systems},
Cambridge University Press, Cambridge (2005).

\bibitem{TT93}
T.~Takatsuka and R.~Tamagaki, \emph{Prog. Theor. Phys. Suppl.}
 No. 112, 27 (1993).


\bibitem{Dean03}
D.~J.~Dean and M.~Hjorth-Jensen,
\emph{Rev. Mod. Phys.} {\bf 75}, 607 (2003).

\bibitem{Lombardo-Schulze}
U.~Lombardo and H.-J. Schulze, In
\emph{Lecture Notes in Physics}, Vol.578, p.30, Springer (2001). 


\bibitem{Schwenk}
A.~Schwenk, B.~Friman, G.~E.~Brown,
\emph{Nucl. Phys.} {\bf A713}, 191 (2003).

\bibitem{Cao-Lombardo} 
L.~G.~Cao, U.~Lombardo, and P.~Schuck,
\emph{Phys. Rev. C} {\bf 74}, 064301 (2006).


\bibitem{Fabrocini05}
A.~Fabrocini, S.~Fantoni, A.~Y.~Illarionov, and K.~E.~Schmidt,
\emph{Phys. Rev. Lett.} {\bf 95}, 192501 (2005).

\bibitem{AbeSeki09}
T.~Abe and R.~Seki, 
\emph{Phys. Rev. C} {\bf 79}, 054002 (2009).

\bibitem{GezerlisCarlson}
A.~Gezerlis and J.~Carlson,
\emph{Phys. Rev. C} {\bf 81}, 025803 (2010).

\bibitem{Gandolfi} 
S.~Gandolfi, A.~Yu.~Illarionov, F.~Pederiva, K.~E.~Schmidt, S.~Fantoni,
\emph{Phys. Rev. C} {\bf 80}, 045802 (2009).

\bibitem{Matsuo06}
M.~Matsuo,
\emph{Phys. Rev. C} {\bf 73}, 044309 (2006).

\bibitem{Tamagaki}

R.~Tamagaki, 
\emph{Prog. Theor. Phys.} {\bf 39}, 91 (1968).

\bibitem{Leggett}
A.~J.~Leggett, 
In eds.  A.~Pekalski and R.~Przystawa
\emph{ Modern Trends in the Theory of
Condensed Matter}, 
Lecture Note in Physics 115,
Springer-Verlag, Berlin, (1980);
A.~J.~Leggett, \emph{J. de Phys.} {\bf 41}, C7-19 (1980).
 
 \bibitem{Nozieres}
P.~Nozi\`{e}res and S.~Schmitt-Rink, 
\emph{J. Low Temp. Phys.}
{\bf 59}, 195 (1985). 

\bibitem{Melo}
C.~A.~R.~S\'{a}~de~Melo, M.~Randeria, and J.~R.~Engelbrecht,
\emph{Phys. Rev. Lett.} {\bf 71}, 3202 (1993).

\bibitem{Engelbrecht}
J.~R.~Engelbrecht, M.~Randeria, and  C.~A.~R.~S\'{a}~de~Melo,
\emph{Phys. Rev. B} {\bf 55}, 15153 (1997).

\bibitem{Randeria}
M. Randeria, In eds. A.~Griffin, D.~Snoke, and S.~Stringari,
\emph{Bose-Einstein Condensation},
Cambridge Univ. Press,  Cambridge, (1995).



\bibitem{Regal}
C.~A.~Regal, M.~Greiner, and D.~S.~Jin,
\emph{Phys. Rev. Lett.} {\bf 92}, 040403 (2004).

\bibitem{review-cold-gas}
S.~Giorgini, L.~V.~Pitaevski, S.~Stringari,
\emph{Rev. Mod. Phys.} {\bf 80}, 1215 (2008).



\bibitem{Hansen}
P.~G.~Hansen and B.~Jonson, 
\emph{Europhys. Lett.} {\bf 4}, 409 (1987).

\bibitem{Esbensen}
G.~F.~Bertsch and H.~Esbensen, \emph{Ann. Phys.} \textbf{209},  327 (1991);
H.~Esbensen and G.~F.~Bertsch, \emph{Nucl. Phys.} \textbf{A542}, 310 (1992).


\bibitem{Ikeda} 
K.~Ikeda, \emph{Nucl. Phys.} {\bf A538}, 355c (1992).

\bibitem{Zhukov}
M.~V.~Zhukov, B.~V.~Danilin, D.~V.~Fedorov, J.~M.~Bang,
I.~J.~Thompson, and J.~S.~Vaagen,
\emph{Phys. Rep.} {\bf 231}, 151 (1993).

\bibitem{Barranco01}
F.~Barranco, P.~F.~Bortignon, R.~A.~Broglia, G.~Col\'{o}, and E.~Vigezzi,
\emph{Eur. Phys. J.} {\bf A11}, 385 (2001).



\bibitem{Hagino05} %thee body model + dipole
K.~Hagino and H.~Sagawa, 
\emph{Phys. Rev. C} {\bf 72}, 044321 (2005).

\bibitem{Hagino07} %BEC-like
 K.~Hagino, H.~Sagawa, J.~Carbonell, and P.~Schuck, 
\emph{Phys. Rev. Lett.} \textbf{99},  022506 (2007).

\bibitem{HaginoIOP} % Cooper pair size in 11Li HaginoIOP
 K.~Hagino, H.~Sagawa, and P.~Schuck,
\emph{J. Phys. G} {\bf 37}, 064040 (2010).

\bibitem{Myo08}  
T.~Myo, Y.~Kikuchi, K.~Kat\={o}, H.~Toki and K.~Ikeda, 
\emph{Prog. Theor. Phys.} 
{\bf 119}, 561 (2008).

\bibitem{Bertsch} 
G.~F.~Bertsch, R.~A.~Broglia, and C.~Riedel,
\emph{Nucl. Phys.} {\bf A91}, 123 (1967).

\bibitem{Ibarra}
R.~H.~Ibarra, N.~Austern, M.~Vallieres, and D.~H.~Feng,
\emph{Nucl. Phys.} {\bf A288}, 397 (1977).

\bibitem{Janouch}
F.~A.~Janouch and R.~J.~Liotta, 
\emph{Phys. Rev. C} {\bf 27}, 896 (1983).

\bibitem{Catara84}
F.~Catara, A.~Insolia, E.~Maglione, and A.~Vitturi,
\emph{Phys. Rev. C} {\bf 29}, 1091 (1984).

\bibitem{Ferreira}
L.~Ferreira, R.~Liotta, C.~H.~Dasso, R.~A.~Broglia, and A.~Winther,
\emph{Nucl. Phys.} {\bf A426}, 276 (1984).

\bibitem{MMS05}
M.~Matsuo, K.~Mizuyama, and Y.~Serizawa,
\emph{Phys. Rev. C} \textbf{71},  064326 (2005).

\bibitem{Matsuo10}
M.~Matsuo and Y.~Serizawa, 
\emph{Phys. Rev. C} {\bf 82}, 024318 (2010).


\bibitem{Matsuo07}
M.~Matsuo, Y.~Serizawa, and K.~Mizuyama,
\emph{Nucl. Phys.} \textbf{A788}, 307c (2007).

\bibitem{Pillet07} 
N. Pillet, N. Sandulescu, and P. Schuck, 
\emph{Phys. Rev. C} \textbf{76}, 024310 (2007).

\bibitem{Pillet10} 
N. Pillet, N. Sandulescu, P. Schuck, and J.~-F.~Berger,
\emph{Phys. Rev. C} \textbf{81}, 034307 (2010).

\bibitem{Mottelson} 
B.~Mottelson, 
In eds.
H.~Nifenecker, J.-P.~Blaizot, G.~F.~Bertsch, W.~Weise,
and F.~David, 
\emph{ Trends in Nuclear Physics, 100 Years
Later}, (Les Houches, Session LXVI, 1996), Elsevier, (1998).

\bibitem{Ring-Schuck}
P.~Ring and P.~Schuck, 
\emph{The Nuclear Many-Body Problem},
Springer-Verlag, (1980).





\bibitem{Nakamura} T.~Nakamura  {\it et al.},
%A.~M.~Vinodkumar, T.~Sugimoto, N.~Aoi,
% H. Baba, D. Bazin, N. Fukuda, T. Gomi, H. Hasegawa, N. Imai, M. Ishihara,
% T. Kobayashi, Y. Kondo, T. Kubo, M. Miura, T. Motobayashi, H. Otsu,
%A. Saito, H. Sakurai, S. Shimoura, K. Watanabe, Y. X. Watanabe,
%T. Yakushiji, Y. Yanagisawa, and K. Yoneda,
\emph{Phys. Rev. Lett.} {\bf 96}, 252502 (2006).

\bibitem{Serizawa09} 
Y.~Serizawa and M.~Matsuo, 
\emph{Prog. Theor. Phys.} \textbf{121}, 97 (2009).

\bibitem{Shimoyama11} H.~Shimoyama and M.~Matsuo, 
\emph{ Phys. Rev. C}
{\bf 84}, 044317 (2011).


\bibitem{HaginoNN} 
K.~Hagino, H.~Sagawa, T.~Nakamura, and S.~Shimoura, 
\emph{Phys. Rev. C} {\bf 80}, 032301(R) (2009).


\bibitem{KikuchiNN} 
Y.~Kikuchi, K.~Kato, T.~Myo, M.~Takashina, and K.~Ikeda,
\emph{Phys. Rev. C} {\bf 81}, 044308 (2010).


\bibitem{NakamuraNN} 
T.~Nakamura, private communication

\bibitem{ChulkovNN} % SimonNN
L.~V.~Chulkov, et al., 
\emph{Nucl. Phys.} {\bf A759}, 23 (2005).


\bibitem{Yoshida62}%pair transfer for BCS gr.st
S.~Yoshida, 
\emph{Nucl. Phys.} {\bf 33}, 685 (1962).

\bibitem{Broglia73}
R.~A.~Broglia, O.~Hansen, and C.~Riedel, In eds.
M. Baranger and E. Vogt,
\emph{ Advances in Nuclear Physics} vol.6, pp.287-457,
Plenum, New York, (1973) 


\bibitem{Glendenning} 
N.~K.~Glendenning, \emph{ Direct Nuclear Reactions},
Academic Press, (1983).


\bibitem{DobHFB2}
J.~Dobaczewski, W.~Nazarewicz, T.~R.~Werner, J.~F.~Berger, C.~R.~Chinn, 
and J.~Decharg\'e, 
\emph{Phys. Rev. C} {\bf 53}, 2809 (1996).



\bibitem{Tanihata08}
I.~Tanihata {\it et al.},
%M.Alcorta, D.Bandyopadhyay, R.Bieri, L.Buchmann, B.Davids,
%	N.Galinski, D.Howell, W.Mills, S.Mythili, R.Openshaw,
%	E.Padilla-Rodal, G.Ruprecht, G.Sheffer, A.C.Shotter, M.Trinczek,
%	P.Walden, H.Savajols, T.Roger, M.Caamano, W.Mittig,
%	P.Roussel-Chomaz, R.Kanungo, A.Gallant, M.Notani, G.Savard,
%	I.J.Thompson,
\emph{Phys. Rev. Lett.} {\bf 100}, 192502 (2008).

\bibitem{Potel}
G.~Potel, F.~Brancco, E.~Vigezzi, and R.~A.~Broglia,
\emph{Phys. Rev. Lett.}  {\bf 105}, 172502 (2010).





\end{thebibliography}
\end{document}